\documentstyle[twoside,fleqn,espcrc2,epsf,rotate]{article}

\newcommand{\AmS}{{\protect\the\textfont2
  A\kern-.1667em\lower.5ex\hbox{M}\kern-.125emS}}

\title{Charm Physics $1996$ -- A Retrospective}

\author{E. Golowich\address{Department of Physics and Astronomy, 
        University of Massachusetts, \\ 
        Amherst, MA 01003 USA}%
        \thanks{Acknowledgement is given the National 
Science Foundation for their support.}}

\begin{document}

\begin{abstract}
A pedagogically oriented review is given of progress made over 
the past year in our understanding of physics related to the 
charm quark.  Included are discussions of the $R_c$ deficit, the $\psi'$ 
anomaly, charm spectroscopy, $D$ nonleptonic decays, searches 
for flavor-changing neutral currents, new limits on 
$D^0 - {\bar D}^0$ mixing and prospects for future 
experimental studies of the charm sector.  
\end{abstract}

\maketitle

\section{INTRODUCTION}

Compared to the publicity given in recent years to the physics of 
both $b$-quarks and $t$-quarks, studies of the $c$-quark might 
appear to have become somewhat of a neglected subject.  Nonetheless, 
the year $1996$ has been a productive one in the study of 
charm physics.  In the following we provide 
a personal slice through the corpus of charm-related material.  
Space limitations do not permit a fully comprehensive review 
of the subject, and we apologize to those whose 
work is unmentioned.  
\section{PRODUCING CHARM}
In order to study properties of charm hadrons, charm must first 
be produced.  The production occurs via scattering (using hadron, 
photon and lepton beams) and decay (from the $Z^0$ boson and 
$B$ hadrons).  

We begin by reviewing two items 
which seemed to garner the lion's share of attention to 
charm physics at this year's conferences.  Each involves 
the creation of $c{\bar c}$ pairs which then evolve into 
hadronic final states.  
\subsection{$R_c$ Crisis}
The $R_c$ `crisis' was the concern that 
the measured ratio $R_c \equiv \Gamma_{Z^0 \to c{\bar c}}/\Gamma_{Z^0 
\to {\rm hadrons}}$ is smaller than the Standard Model (hereafter SM) 
value, $R_c^{\rm SM} = 0.1725$.  As recently as 
Moriond '96,$^{\cite{Mor96}}$ one had $R_c^{\rm expt} = 0.1598(69)$ 
(a discrepancy of $-1.8~\sigma$), yet by DPF96$^{\cite{Dpf96}}$ 
the value had changed to $R_c^{\rm expt} = 0.1715(56)$ (a discrepancy 
of $-0.1~\sigma$).  

Thus the $R_c$ crisis is no more.  Together with an enlarged data 
sample, it was imposition of improved experimental technique 
which led to the upward revision, including double-tagging procedures 
and improved verification of closure.  Regarding the latter, 
we mean by `closure' the assumption that all $c{\bar c}$ pairs 
produced at the $Z^0 \to c{\bar c}$ vertex ultimately 
appear as charm hadrons ($D^0, D^+, D_s^+, \Lambda_c^+, \dots$) 
which then decay weakly.  Any correct determination of $R_c^{\rm expt}$ 
must account for all such decay products, and in fact, a 
careful channel-by-channel analysis of final states such as 
$K^-\pi^+~(D^0)$, $K^-\pi^+\pi^+~(D^+)$, $\phi^0\pi^+~(D_s^+)$ and 
$p K^-\pi^+~(\Lambda_c^+)$ succeeded in accurately accounting for 
all decay products from charm.  Moreover, a change in the assumed 
$D^0 \to K^-\pi^+$ branching ratio down to its current value also 
contributed to the rise in $R_c^{\rm expt}$.  
\subsection{$\psi'$ Anomaly at CDF}
This refers to the program undertaken by CDF to 
study $p{\bar p} \to \psi'(3686) + X$ at transverse 
momenta as large as $p_T = 20$~GeV.$^{\cite{Schm96},\cite{Ka96}}$ 
The history of theoretical attempts to analyze charmonium production 
follows a somewhat indirect path.  The original expectation was that 
inclusive $B$ decay ($B \to \psi' + X$) would give rise to the $\psi'$
yield.  Yet only about $23\%$ could be so identified, the majority 
having a `prompt' origin.  The leading order (LO) gluon-fusion
mechanism ($gg\to c{\bar c}[^3S_1] + g$) for prompt production
predicted $d\sigma/dp_T$ to be too small and with incorrect $p_T$ 
dependence.  Although it became appreciated that $\psi'$ production 
at large $p_T$ is dominated by fragmentation of the $c$-quark and 
gluon jets, the normalization for production of color-singlet 
$c{\bar c}$ pairs was still too small.  Finally, including both 
color-singlet and color-octet $c{\bar c}$ pairs$^{\cite{BrFl94}}$ 
yielded agreement with the data, although at the cost of fitting 
certain parameters associated with nonperturbative 
contributions.$^{\cite{BoBrLe95}}$  The fits to $p{\bar p} \to 
\psi' + X$, however, do not work when applied to 
$\gamma p \to J/\psi + X$.$^{\cite{CaKr96}}$   

It is clear that further theoretical and experimental work 
is yet required.  The theory underlying charmonium production 
is subtle and difficult, and will take some time to sort out.  
A review of the basics appears in Ref.~\cite{BrFlYu96} and other 
analyses appear regularly.$^{\cite{new96}}$  Further experimental 
tests are needed, like that involving the prediction of 
$\psi'$ transverse polarization$^{\cite{Wi95}}$ 
associated with the color-octet mechanism.  
\subsection{Additional topics}
Charm production is a large subfield of charm physics, 
and the following four topics are chosen to illustrate 
the range of activity:

(a) Scattering experiments continue to test the body of QCD predictions 
for charm hadron production.  Thus, Fermilab experiment E769 
examined charm production off a nuclear target using $\pi$,~$K$,~$\rho$ 
projectiles.$^{\cite{Alv96}}$   The energy dependence for forward 
production of several types of charm hadrons ($D^0, \Lambda_c^+ ,
\dots$) was found to agree with that predicted by perturbative QCD, 
and distributions of $x_F$ and $p_T^2$ were measured for $D$-meson 
production as a probe of gluon distributions in the target and 
beam particles.  

(b) It is important to have proper theoretical understanding of 
heavy-quark ({\it e.g.} charm) contributions to deep inelastic 
structure functions over the very different kinematical regions 
$Q^2 \sim m_c^2$ and $Q^2 \gg m_c^2$ in order to correctly evolve 
parton distributions in $Q^2$ as heavy-quark thresholds are
encountered. A recent theoretical analysis is given in Ref.~\cite{Ma96}. 

(c) Work appears on charm production in models of new physics. 
In the class of examples having two Higgs doublets, 
flavor-changing couplings exist even at tree level and a decay such 
as $t \to c ~\gamma$ becomes possible.$^{\cite{AtReSo96}}$    

(d) Finally, attention has been directed over the past 
several years at the number of $c$-quarks produced per $B$ decay 
and its possible relation to the discrepancy between theory 
and experiment regarding the $B$ semileptonic branching ratio.  
For a summary given earlier this year, see Ref.~\cite{BrHo96}. 
\section{CHARM HADRON PROPERTIES}
From the viewpoint of performing accurate predictions of the SM, 
the $c$-quark mass scale presents a nontrivial challenge.  
Methods of chiral symmetry are not generally applicable because 
$m_c$ is too large, and the use of heavy-quark 
methodology (which incorporates both heavy-quark effective 
theory (HQET) and heavy-quark expansions based on the operator product
expansion (OPE)) is questionable because $m_c$ might not be large
enough.  It will, of course, take more time and effort to decide on 
the latter point, although the large number of resonances 
in the charm region (as cited in the $1996$ Particle Data Group 
listing$^{\cite{Pd96}}$) reflects the vigor of QCD dynamical 
activity and makes problematic the assumption of local 
duality. The literature, reflecting the lack of a clear winner in the 
charm region, contains quark models and heavy-quark methods alike, 
often used in conjunction with additional approaches like 
lattice-QCD, the $1/N_c$ expansion and QCD sum rules.

Work continues, however, and there are a number of very active 
areas involving the physics of charm hadrons. Among these are 
spectroscopy, weak decays, lifetimes and charmonium studies.  
In view of space limitations, I defer to the summary 
on the final item of Ref.~\cite{Cha96}. 
\subsection{Charm Spectroscopy}
Although it may not yet be clear to what extent heavy-quark methods 
yield reliable quantitative SM predictions in the 
charm sector, it seems to me beyond doubt that they are already 
indispensable as an organizing principle.  This is nowhere more 
clear than in charm spectroscopy.
\subsubsection{Charm Mesons}
A simple example occurs with the 
$D$ meson ground-state and first-excited states.  
Recall some elementary bookkeeping.  For a ($Q~{\bar q}$) meson 
in which the heavy quark has spin ${\bf s}_Q$ and the light quark 
has spin ${\bf s}$ and orbital angular momentum $\ell$, 
the meson spin ${\bf S}$ is found via the chain 
\begin{equation}
{\bf j_\ell} = {\bf s} + {\bf \ell}  \quad 
\Longrightarrow \quad {\bf S} = {\bf j_\ell} + {\bf s}_Q \ \ .
\end{equation}
The pattern of states thus obtained in the $D$ system is 
compiled in Table~{\ref{tab:dspect}. 

\begin{table}[hbt]
\caption{Angular momentum coupling for $Q{\bar q}$.}
\label{tab:dspect}
\begin{tabular}{c|c|c|c|c|c}
\hline
${\bf \ell}$ & $0$ & \multicolumn{2}{c|}{$1$}
&\multicolumn{2}{c}{$2$} \\ 
\hline\hline
${\bf j_\ell}$ & $1/2$ & $1/2$ & $3/2$ & $3/2$ & $5/2$ \\
${\bf S}$ & $0,1$ & $0,1$ & $1,2$ & $1,2$ & $2,3$ \\ 
\hline
\end{tabular}
\end{table}

There will be transitions between these states, 
and these are displayed in Figure~{\ref{fig:dmesons}.  The 
selection rules obtained from angular momentum and 
parity conservation are listed in Table~\ref{tab:sr}.

\begin{table}[hbt]
\caption{Angular momentum and parity selection rules.}
\label{tab:sr}
\begin{tabular}{cccc}
\hline
$D_2^* \to D$ & $D_1 \to D$ & $D_1 \to D^*$ & $D_2^* \to D^*$ \\ 
\hline\hline
$2$ & --- & $0,2$ & $2$ \\ 
\hline
\end{tabular}
\end{table}

The selection rules explain why there is no $D_1 \to D$ transition in 
Figure~{\ref{fig:dmesons}.  Invoking heavy-quark symmetry at this 
point yields an important constraint not covered by the 
above selection rules, that the $S$-wave 
$D_1 \to D^*$ amplitude vanishes, so to leading order 
all three transition amplitudes are $D$-wave.  According to 
Ref.~{\cite{Fa96a}, heavy-quark symmetry can reach quantitative 
agreement with the observed decay rates provided the leading-order 
predictions are corrected by ${\cal O}(m_c^{-1})$ effects.  
\begin{figure}[htb]
\vspace{9pt}
%\framebox[55mm]{\rule[-21mm]{0mm}{43mm}}
\epsfbox{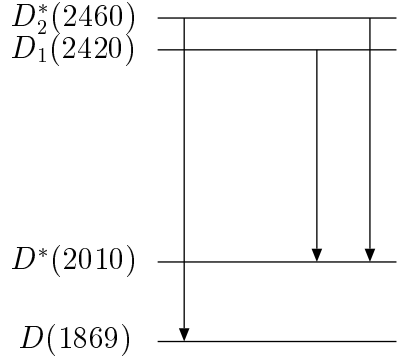}
\caption{$D$ meson transitions.}
\label{fig:dmesons}
\end{figure}
\subsubsection{Charm Baryons}
The spin of a $(Q\ q_1~q_2)$ baryon having heavy-quark spin 
${\bf s_Q}$ and spectator angular momenta ${\bf s}_{12}, 
{\bf \ell}$ is constructed as 
\begin{equation}
{\bf j_\ell} = {\bf s}_{12} + {\bf \ell}  \quad 
\Longrightarrow \quad {\bf J} = {\bf j_\ell} + {\bf s}_Q 
\end{equation}
One thus obtains the nine ground state baryons, partitioned into a flavor 
antitriplet and a flavor sextet, shown in Table~\ref{tab:bary}.  

\begin{table}[hbt]
\caption{Baryon Ground State}
\label{tab:bary}
\begin{tabular}{cccc}
\hline 
Color & Flavor & $s_{12}$ & $J^P$ \\ \hline\hline
${\bf 3^*}$ & ${\bf 3^*}$ & $0$ & $(1/2)^+$ ($\Lambda_c, \Xi_c$)\\
\hline
${\bf 3^*}$ & ${\bf 6}$ & $1$ & $(1/2)^+$ ($\Sigma_c, \Xi_c',\Omega_c$) \\
 & & & $(3/2)^+$ ($\Sigma_c^*, \Xi_c^*,\Omega_c^*$) \\
\hline
\end{tabular}
\end{table}

It is traditional that data on baryon spectroscopy lags 
somewhat behind that of meson spectroscopy, and such is the case 
for the charm sector.  However, our knowledge of the charm 
baryon ground state was enhanced in 1996 by contributions 
from the CLEO collaboration, which cited evidence for 
$\Xi_c^{*+}(2645)$$^{\cite{Cl96a}}$ and for 
$\Sigma^{*++}(2520)$, $\Sigma^{*0}(2518)$$^{\cite{Cl96b}}$.  
Guided by earlier model calculations$^{\cite{Jrms95}}$, 
each was assigned as a $S = 3/2$ ground state charm baryon.  

If one accepts the CLEO quantum number assignments, this 
leaves $\Xi_c'$ and $\Omega_c^*$ to be discovered.  Jenkins 
used mass sum rules derived from an expansion in $1/m_c$, $1/N$ 
and $SU(3)$ breaking together with input mass values of $\Lambda_c, 
\Xi_c, \Sigma_c, \Omega_c, \Sigma_c^*, \Xi_c^*$ to 
predict $m_{\Xi_c'} = 2580.8 \pm 2.1$ and 
$m_{\Omega_c^*} = 2760.5 \pm 4.9$.$^{\cite{Je96}}$   
Falk, on the other hand, used the evident violation of a sum 
rule based on heavy-quark and $SU(3)$ relations, 
\begin{equation}
0.84 \simeq {m_{\Sigma_b^*} - m_{\Sigma_b} \over m_{\Sigma_c^*} 
- m_{\Sigma_c}} 
= {m_{B^*} - m_B \over m_{D^*} - m_D} \simeq 0.33 \ \ ,
\end{equation}
to cast doubt on the the existing $\Sigma_c$ and $\Sigma_c^*$ 
assignments.$^{\cite{Fa96b}}$  Although it might be the 
$\Sigma_b^* - \Sigma_b$ mass difference which causes the problem, 
the need for direct spin-parity assignments is good 
to keep in mind.  Finally, the subject of mixing between baryons 
in the flavor ${\bf 6}$ and ${\bf 3}^*$ multiplets 
was analyzed in Ref.~\cite{It96} and in Ref.~\cite{Sa96}. 

\subsection{Charm Lifetimes} 
Another area for application of heavy-quark
expansions$^{\cite{Sh95},\cite{Ur96}}$ is the set of charm lifetimes.  
At the most elementary level, the decay rate for a heavy hadron 
$H_Q$ containing a single heavy quark $m_Q$ can be expressed as 
\begin{equation}
\Gamma (H_Q) = \Gamma_\infty + \Gamma_{\rm non-univ} \ \ .
\end{equation}
The term $\Gamma_\infty$ is universal among the set of hadrons 
$\{ H_Q \}$ and dominates for $m_Q /\Lambda_{\rm QCD} \gg 1$.  
It is as if the heavy quark were free.  The differences in 
individual $H_Q$ lifetimes arise in $\Gamma_{\rm non-univ}$ 
from various nonuniversal effects. The display of charm hadron 
lifetimes in Figure~\ref{fig:life}, most dramatically the ratio 
\begin{equation} 
{\tau_{D^+} \over \tau_{\Omega_C^0}} \ = \ 16.5 \pm 5.2 \ \ ,
\end{equation}
reveals that such terms must play a significant role for charm lifetimes.  
\begin{figure}[htb]
%\framebox[55mm]{\rule[-21mm]{0mm}{43mm}}
\epsfbox{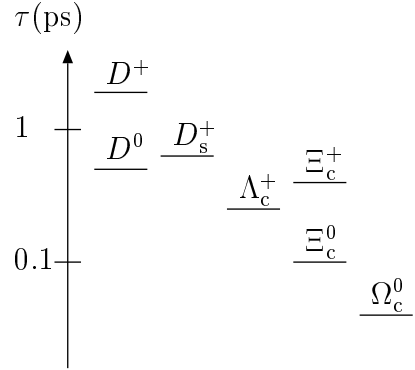}
\caption{Lifetimes of the charm hadrons.}
\label{fig:life}
\end{figure}
In the heavy-quark approach, the decay rate is expressed 
as an expansion in inverse powers of $m_Q$, 
\begin{eqnarray}
& & \Gamma_{H_Q\to f} = {G_F^2 m_Q^5 \over 192\pi^3} |{\rm KM}|^2 
\nonumber \\ 
& & \phantom{xxxx}\times \left[ A_0 + {A_2 \over m_Q^2} + 
{A_3 \over m_Q^3}  + {\cal O}(m_Q^{-4}) \right] \ ,
\end{eqnarray}
where $|{\rm KM}|^2$ gives the CKM dependence of the weak decay. 
Each of the leading $\{A_n\}$ can be interpreted physically.  
Thus for charm decay, $A_2$ contains dependence on (i) the kinetic 
energy of the $c$-quark and (ii) the spin-spin interaction between the 
$c$-quark and the spectators, whereas $A_3$ includes (iii) weak interaction 
effects like $c \to su{\bar d}$, $c{\bar q} \to q_1{\bar q}_2$, 
$cq \to q_1q_2$ as well as (iv) a Pauli interference effect 
which occurs if a spectator quark in the initial state is identical 
to a quark which appears in a final state in one of the weak
transitions of (iii).  

Application of this approach to the charm lifetimes is reasonably 
successful.  There appear to be no major disasters, 
but model dependence remains in estimates of the $\{A_n\}$.  
We recommend the recent summary by Bigi$^{\cite{Bi96}}$. 
\subsection{Leptonic and Semileptonic Decays}
The leptonic and semileptonic charm decays are reviewed in 
Ref.~\cite{BuRi95}.  Charm leptonic decays 
are important because they determine the decay constants 
$f_D$ and $f_{D_s}$.  There are 
now several direct measurements of $f_{D_s}$, including 
the recent E653 value $f_{D_s} = 194 \pm 35 \pm 20 \pm 14$ 
MeV.$^{\cite{Ko96}}$  Lattice-theoretic values of charm decay 
constants presented at LATTICE96 are similar to those of previous 
years, {\it e.g.} the JLQCD collaboration$^{\cite{JLQCD96}}$ cites 
$f_D = 202(8)^{+24}_{-11}$ MeV and $f_{D_s} = 216(6)^{+22}_{-15}$ 
whereas the MILC collaboration$^{\cite{MILC96}}$ has 
$f_D = 196(9)(14)(8)$ MeV and $f_{D_s} = 211(7)(25)(11)$.   

In an interesting contribution to the literature of charm 
semileptonic transitions, Voloshin argues that large 
effects are to be expected from Pauli interference of the 
$s$-quark in $\Xi_c$ and $\Omega_c$ semileptonic decay, 
implying sharply enhanced decay rates relative to 
$\Lambda_c$.$^{\cite{Vo96}}$ 
\subsection{Nonleptonic Weak Decays}
The data base for nonleptonic decays continues to expand, and 
a review of the current situation appears in Ref.~\cite{BrHo96}.  
As an example, let me cite the recent CLEO analysis of $D^0 \to K{\bar K}X$ 
decays.$^{\cite{As96}}$  This analysis covers the five 
final states ($K^+K^-$, $K^0{\bar K}^0$, $3K_{\rm S}$, 
$\pi^0 K_{\rm S}K_{\rm S}$, $K^+K^-\pi^0$), all of which 
have branching ratios well under a per~cent.  Detection of the 
final two modes represents first observations.  

As the study of charm nonleptonic continues, one awaits 
progress in the two-body final state sector, where many modes 
such as $D^+ \to K^+ \eta$ {\it etc.} remain undetected. 
Theorists have the most to say about two-body modes and 
advances of the data set in this area would be welcome. 

The theoretical study of charm nonleptonic decays is 
notoriously difficult, and despite the efforts of many over 
a number of years there does not exist at present a practical 
quantitative approach which follows rigorously from first principles.  
In principle, there seems to be no insurmountable barrier 
to using lattice-theoretic methods.$^{\cite{Ci96}}$ 
Until such time as lattice studies take over, however, it will 
be necessary to proceed in a more traditional manner.  
I shall focus on an interesting contribution which appeared 
from Buccella, Lusignoli and Pugliese (BLP).$^{\cite{Bu96}}$ 
Their approach incorporates the usual collection of 
quark diagrams, the latest renormalization group improved 
weak hamiltonian, and most notably, final-state interaction 
(FSI hereafter) effects.$^{\cite{ClLi96}}$  The work is comprehensive and 
many modes are taken into account.  The authors themselves point out 
some shortcomings, like the assumption 
of factorization, the use of some as-yet unobserved resonances 
to generate the FSI and the many free parameters used (in the latest fit, 
there are $49$ data points, $15$ parameters, with $\chi^2 \simeq 70$).
Not all predictions are successful, such as those for 
the modes $D_s^+ \to \rho^+\eta'$ and $D^0 \to K^{*0} \eta$.  
As a whole, however, the BLP analysis represents a plausible 
theoretical laboratory which (one hopes) provides 
a reasonable picture of the two-body nonleptonic charm decays.  

Two recent contributions of the BLP group 
are of special interest because both involve 
as-yet unobserved signals.  The first concerns a set of predictions 
for CP-violating (CPV hereafter) asymmetries.$^{\cite{Bu95},\cite{Bu96}}$   
We comment on these quantities more fully in Sect.~4.2.4 of this 
report.  It suffices here to note that Ref.~\cite{Bu96} cites as 
the best candidate for detection the asymmetry 
\begin{eqnarray}
a^{\rm CPV}_{(\rho\pi)} &\equiv& {\Gamma_{D^+\to \rho^0\pi^+} 
- \Gamma_{D^- \to \rho^0\pi^-} \over 
\Gamma_{D^+\to \rho^0\pi^+} + \Gamma_{D^- \to \rho^0\pi^-}} \nonumber \\
&\simeq& -2 \times 10^{-3} \ \ .
\end{eqnarray}
More generally, the predicted CPV asymmetries occur at the 
${\cal O}(10^{-3})$ level.  Another BLP result concerns the difference
in decay rates between $D^0$ and ${\bar D}^0$, as obtained from an 
explicit sum over exclusive modes, 
\begin{eqnarray*}
& & {\Delta \Gamma_D \over \Gamma_D} \simeq   2 
{\Gamma_{12}^{(D)} \over \Gamma_D} \simeq (1.5 + i~0.0014)\cdot
10^{-3} \ ,\\
& & \Gamma_{CP = +1} > \Gamma_{CP = -1} \ .
\end{eqnarray*}
To even attempt such an estimate of $\Gamma_{12}^{(D)}$, 
it is first necessary to have a fairly complete collection 
of decay amplitudes (magnitudes {\it and} phases).  
The small magnitude found for $\Delta \Gamma_D /
\Gamma_D$ is noteworthy because large $SU(3)$ breaking effects 
observed in individual decay rates would suggest a rather 
larger value.  On a qualitative level, this reinforces 
the prediction$^{\cite{Ge92}}$ from heavy-quark theory 
that $D^0 - {\bar D}^0$ mixing is smaller than that 
expected from a dispersive approach$^{\cite{DoGoHo86}}$ which 
stresses the large $SU(3)$ breaking.  
\section{FCNC STUDIES}
Due to CKM suppression, the charm sector is not 
the best of places to seek SM signals associated with 
flavor-changing neutral current effects.  
However, this makes charm processes 
attractive for exploring various new 
physics effects.$^{\cite{BuGoHePa97}}$   Below, we first 
review attempts to detect flavor-changing neutral 
current decays and then turn to the subject of 
$D^0 - {\bar D}^0$ mixing.  

\subsection{FCNC Decays}
No flavor-changing neutral current charm decays have yet 
been observed.  We list some upper bounds (at $90\%$~C.L.) 
announced in $1996$: 
\begin{itemize}
  \item E791:$^{\cite{Ai96a}}$ Search for $D^+ \to \pi^+ \ell^+\ell^-$ \\ 
   $B_{D^+ \to \pi^+ e^+e^-} < 6.6\times 10^{-5}$ \\
   $B_{D^+ \to \pi^+ \mu^+\mu^-} < 1.8\times 10^{-5}$ \\
  \item E771:$^{\cite{Al96}}$ Search for $D^0 \to \mu^+ \mu^-$  \\
   $B_{D^0 \to \mu^+\mu^-} < 4.2\times 10^{-6}$ \\ 
  \item CLEO:$^{\cite{Fr96}}$ Search for $D^0 \to \ell^+ 
   \ell^-,~X^0\ell^+ \ell^-$  \\
  $B_{D^0 \to \ell^+\ell^-,~X^0\ell^+ \ell^-} 
   < {\cal O}(10^{-4} \to 10^{-5})$ \\ 
  \end{itemize}

Even if a FCNC $D$-decay is found, one will need to exercise 
some caution when interpreting the result.  For example, 
consider the weak radiative transitions $D\to M\gamma\ (M 
= \rho, K^*, {\it etc})$. The current level of sensitivity is 
$B_{D \to M\gamma} = {\cal O}(10^{-4})$.  
Now, in the absence of QCD radiative corrections, the associated quark
(or short distance) branching ratio is easily 
calculated to be $B^{(0)}_{c \to u\gamma} \simeq 1.4 \times 10^{-17}$.  
Although QCD corrections are found$^{\cite{BuGoHePa96},\cite{Gre96}}$ 
to increase this value appreciably ({\it e.g.} Ref.~\cite{BuGoHePa96} finds 
$B_{c \to u\gamma} \simeq 5 \times 10^{-12}$), the effect remains 
unobservable.  It would seem then that 
observation of $D \to M\gamma$ would signal the presence of new physics.  
However, there are still long range effects to be considered.  
Several such analyses for $B_{D \to M\gamma}$ have been 
carried out in terms of vector dominance$^{\cite{BuGoHePa96}}$ 
and weak annihilation$^{\cite{Kh95}}$ mechanisms, 
with the result $\Longrightarrow$ $B_{D\to M\gamma} = {\cal
O}(10^{-6})$.  See also Ref.~\cite{Si96} The lesson 
is that misinterpretation of FCNC signals is a real 
possibility unless long range effects are first 
understood.$^{\cite{BuGoHePa97}}$   

\begin{figure}[htb]
%\framebox[55mm]{\rule[-21mm]{0mm}{43mm}}
\epsfbox{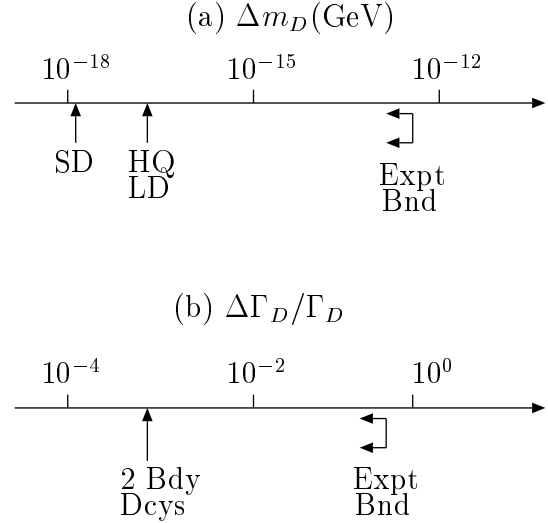}
\caption{Current status of $D^0 - {\bar D}^0$ mixing: 
(a) $\Delta m_D$ in units of GeV, (b) $\Delta \Gamma_D/\Gamma_D$.}
\label{fig:primer}
\end{figure}

\subsection{Mixing}
Of the recent occurrences involving charm, some of 
the most interesting have concerned $D^0 - {\bar D}^0$ mixing.  
We first give a brief review of the subject and then consider 
recent theoretical and experimental developments.  
\subsubsection{Basics of $D^0 - {\bar D}^0$ mixing}
A glance at Figure~\ref{fig:primer} reminds us of the 
key features of this subject.  Neither a difference in mass 
$\Delta m_D$ nor a difference in rate $\Delta \Gamma_D$ has yet been 
observed, 
\begin{eqnarray}
|\Delta m_D| &<& 1.3\times 10^{-13}~{\rm GeV} \ \ , \nonumber \\
|\Delta \Gamma_D| &<& 2.7 \times 10^{-13}~{\rm GeV} \ \ .
\end{eqnarray}
An equivalent set of quantities often more 
useful for direct comparison of theory with experiment, 
together with their bounds is 
\begin{eqnarray}
x_D &\equiv& {|\Delta m_D| \over \Gamma_D} < 0.088 
\ , \nonumber \\
y_D &\equiv& {|\Delta \Gamma_D| \over 2 \Gamma_D} 
< 0.085 \ \ , 
\end{eqnarray}
as well as the parameter 
\begin{equation}
r_f \equiv {B_{D^0 \to {\bar D}^0 \to {\bar f}} \over 
B_{D^0 \to f}} \ \ .
\end{equation}
\begin{figure}[htb]
%\framebox[55mm]{\rule[-21mm]{0mm}{43mm}}
\rotate[l]{
\epsfbox{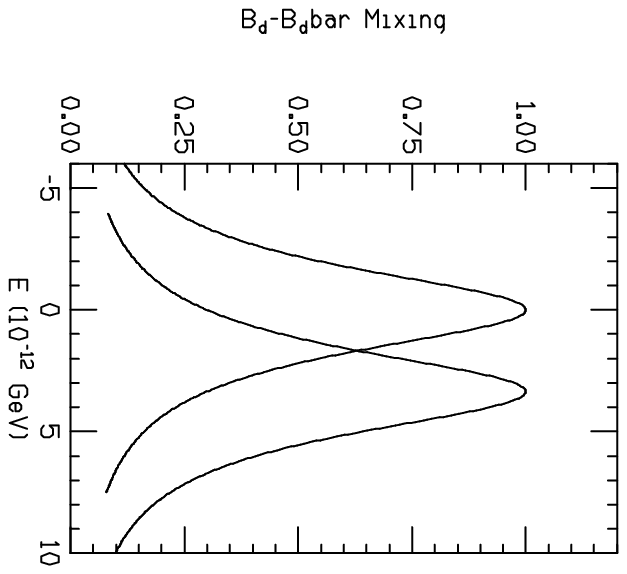}}
\caption{Profile of $B_d - {\bar B}_d$ mixing.} 
\label{fig:bdmix}
\end{figure}
\begin{figure}[htb]
%\framebox[55mm]{\rule[-21mm]{0mm}{43mm}}
\rotate[l]{\epsfbox{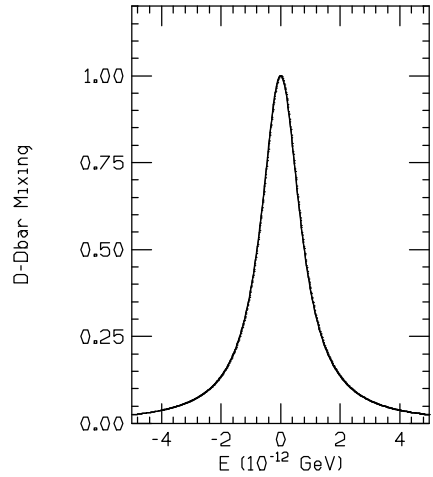}}
\caption{Profile of $D - {\bar D}$ mixing.} 
\label{fig:dmix}
\end{figure}
To get some feeling for the difficulty in detecting 
$D^0 - {\bar D}^0$ mixing, we 
compare the situation for mixing in the $B_d$ system 
(Figure~\ref{fig:bdmix}) with that in the $D$ system 
(Figure~\ref{fig:dmix}).  In these figures we plot 
the imaginary parts of the Breit-Wigner profile for 
each particle.  Thus, in Figure~\ref{fig:bdmix} 
the separation of the two peaks gives $\Delta m_{B_d}$ 
and since the SM expectation is that 
$\Delta m_{B_d} \gg \Delta\Gamma_{B_d}$ the peaks have 
the same widths.  An analogous depiction for the $D$ system 
is given in Figure~\ref{fig:dmix} except it appears that 
one of the peaks has been forgotten.  However, there actually 
are two peaks but they are so closely spaced that in a 
graph where the decay width has a reasonable proportion,  it is not 
possible to distinguish the $D^0$ and ${\bar D}^0$ profiles.  
For an instructive comparison of all four meson mixings, 
see Ref.~\cite{Go95}.

\subsubsection{Theoretical aspects of $D^0 - {\bar D}^0$ mixing}
There exist various theoretical estimates of $|\Delta m_D|$, 
which in Figure~\ref{fig:dmix} we have partitioned into two broad 
categories, short-distance (SD) and 
heavy-quark/long-distance (HQ and LD).  The SD estimate 
arises from the box diagram at the quark 
level,$^{\cite{Da85},\cite{DoGoHo86}}$   
\begin{equation}
|\Delta m_D^{{\rm box}}|  \simeq 5\times 10^{-18}~ {\rm GeV} \ \ .
\end{equation}
The heavy-quark$^{\cite{Ge92}}$ and 
long-distance$^{\cite{DoGoHo86}}$ contributions 
are larger in magnitude.  Although the difference between 
these latter values might be of some theoretical 
interest, the really important point for us is 
that each is far below the current experimental limit. 
With this observation, we are ready to consider a recent 
flurry of activity on the topic of time dependence in $D^0 - {\bar D}^0$ 
nonleptonic mixing amplitudes.

For $x_D, y_D \ll 1$ and $|\lambda| \ll 1$ ($\lambda$ is defined below), 
the time-dependent decay rate for the transition of $D^0$ 
to some final state $f$ time can be written 
as$^{\cite{Li95},\cite{Bl95}}$  
\begin{equation}
\Gamma_{D^0 (t) \to f} \propto e^{-\Gamma_D t}~\bigg[ X + Y~t + Z~t^2
\bigg]\ \ .
\end{equation}
The constant term $X = 4 |\lambda|^2$ arises from doubly-suppressed 
Cabibbo decay (DCSD).  It mimics the mixing signal and is present even 
in the absence of mixing. The quadratic term $Z$ is entirely due to mixing, 
\begin{equation}
Z \equiv (\Delta m_D)^2 + \left( {\Delta \Gamma_D 
\over 2} \right)^2 \ \ .
\end{equation}
The linear term $Y$ arises from interference and 
can be written as 
\begin{equation}
Y = 2 {\cal R}e~\lambda ~\Delta \Gamma_D  + 
   4 {\cal I}m~\lambda ~\Delta m_D  \ \ ,
\end{equation}
where $\lambda$ is a complex number defined by 
\begin{equation}
\lambda \equiv {p \over q}~{A \over B} \ \ ,
\end{equation}
with $p$ and $q$ being the usual mass matrix parameters and 
\begin{equation}
A \equiv \langle f | H_{\rm wk} | D^0 \rangle \ , \qquad 
B \equiv \langle f | H_{\rm wk} | {\bar D}^0 \rangle \ .
\end{equation}
Authors have tended to argue that the $\Delta\Gamma_D$ contribution to 
the $Y$ term is negligible.  The remaining contribution 
to $Y$ is proportional to ${\cal I}m~\lambda$, which 
can be nonzero if (i) CPV is present (thus inducing a phase in $p/q$) 
and/or (ii) the FSI are different in $D^0 \to f$ and ${\bar D}^0 
\to f$ (thus inducing a phase in $A/B$).  

Two papers appeared in 1995 each pointing out the potential 
importance of the $Y$-term for studies of mixing in the 
$D$ sector.  Wolfenstein argued that detection of mixing 
at current levels of sensitivity would require new physics, 
and that in the absence of FSI, the effect would 
involve CPV.$^{\cite{Wo95}}$   Blaylock {\it et al} 
performed a careful analysis of the time dependence in mixing 
and discussed the importance of allowing for FSI effects in any 
model-independent experimental analysis.$^{\cite{Bl95}}$   
More recently, Browder and Pakvasa provided a more detailed 
look at new physics possibilities.$^{\cite{Br96}}$ They also 
stressed that the term $Y$ would survive (i) for zero CPV 
in the combination $\Gamma_{D^0 (t) + {\bar D}^0 (t)}$ and 
(ii) for zero FSI in the combination $\Gamma_{D^0 (t) - 
{\bar D}^0 (t)}$.  CPV contributions to $D^0 - {\bar D}^0$ mixing 
from a variety of new physics scenaria are summarized in 
Ref.~\cite{BuGoHePa95} and in Ref.~\cite{Lo96}.  
\subsubsection{Results of E791}
A number of results from the Fermilab experiment E791 appeared 
during the past year.  This experiment recorded $2 \times 10^{10}$ 
raw events from $500$~GeV/c $\pi^-$ interactions, and the 
ultimate yield of about $2 \times 10^{5}$ reconstructed charm 
is the largest sample to date.  No signal for $D^0 - {\bar D}^0$
mixing was found by E791.  In the following we list some specific 
results, beginning in Table~\ref{tab:semimix} with a bound on mixing 
obtained from semileptonic decay$^{\cite{Ai96b}}$ and including 
for comparison the earlier E615 result.  

\begin{table}[hbt]
\caption{Mixing limits at $90\%$~C.L. (semileptonic).}
\label{tab:semimix}
\begin{tabular}{lcc}
\hline
Experiment & E615 & E791  \\ \hline\hline
$r_{K\ell{\bar \nu}}^{\rm mix}$ & $< 0.56~\%$ & $< 0.50~\%$ \\ 
\hline
\end{tabular}
\end{table}

Upon assuming zero mixing 
in its nonleptonic data sample, E791 obtained$^{\cite{Ai96c}}$ 
a DCSD signal in the $K\pi$ mode comparable to the wrong-sign signal 
obtained earlier by CLEO, as shown in Table~\ref{tab:dcsd}. 

\begin{table}[hbt]
\caption{DCSD (assuming $r_D^{\rm mix} =0$).}
\label{tab:dcsd}
\begin{tabular}{lc}
\hline
Experiment & $r_D^{\rm DCSD}$ \\ \hline\hline
CLEO  & $(0.77 \pm 0.25 \pm 0.25)~\%$ \\
E791  & $(0.68 {}^{\ + 0.34}_{\ -0.33} \pm 0.07)~\%$ \\
\hline
\end{tabular}
\end{table}

Finally, Table~\ref{tab:nlmix} displays the mixing values 
obtained from nonleptonic decays ($D \to K\pi$ and $D\to K 3\pi$) 
and obtained under the most general of conditions, 
{\it i.e.} making no assumptions regarding the 
absence of CPV.  The $90\%$~C.L. upper bounds are 
$r_{K\pi,K3\pi}^{\rm mix}({\bar D}^0 \to D^0) < 0.74\%$ and 
$r_{K\pi,K3\pi}^{\rm mix}(D^0 \to {\bar D}^0) < 1.45\%$.   

\begin{table}[hbt]
\caption{Mixing signals (nonleptonic).}
\label{tab:nlmix}
\begin{tabular}{lc}
\hline
Mixing & $r_{K\pi,K3\pi}^{\rm mix}$     \\ \hline\hline
${\bar D}^0 \to D^0$ & $(0.18^{\ +0.43}_{\ -0.39} \pm 0.17)\%$ \\ 
$D^0 \to {\bar D}^0$ & $(0.70^{\ +0.58}_{\ -0.53} \pm 0.18) \%$ \\ 
\hline
\end{tabular}
\end{table}

The reader might wonder, despite the large number of events that E791 
generated, why its bounds are not even lower.  It is important 
to keep in mind that E791 was a hadroproduction fixed-target 
experiment.  In this setting, virtually all charm particles produced 
were highly energetic ($p_D \ge 20$~GeV) and forwardly 
directed ($\theta \le 20$~mr) with a large multiplicity of charged 
particles ($n_{\rm ch} \gg 1$).   This presents a difficult
environment for extracting a clean mixing signal.  
\subsubsection{Bounds on CP violation}
Lastly, we report on the E791 search for CPV in charged 
$D$ decays.$^{\cite{Ai96d}}$  Studies of CPV 
can be divided into two categories, indirect or 
direct.$^{\cite{DoGoHo92}}$  The former consists of 
mixing-induced CPV and occurs only for neutral mesons.  
It is the latter we consider here, for which FSI constitute 
a necessary ingredient.  Because the $c$-quark mass scale is 
still in the resonance region, FSI are clearly present and 
presumably large.  One defines a CPV asymmetry associated 
with decays to a final state $f$ and its conjugate ${\bar f}$ as 
\begin{equation}
a^{\rm CPV}_f \equiv {\Gamma_{D\to f} - \Gamma_{{\bar D} \to {\bar f}}
\over \Gamma_{D\to f} + \Gamma_{{\bar D} \to {\bar f}}} \ \ .
\end{equation}
The E791 results are compiled in Table~\ref{tab:asy}.

\begin{table}[hbt]
\caption{CPV asymmetries in $D^\pm$ decay.}
\label{tab:asy}
\begin{tabular}{lc}
\hline
Mode ($f$) & Asymmetry \\ \hline\hline
$K^-K^+\pi^+$ & $-0.014 \pm 0.029$ \\ 
$\phi\pi^+$ & $-0.028 \pm 0.036$ \\ 
${\bar K}^{*0}(892)K^+$ & $-0.010 \pm 0.050$ \\
$\pi^-\pi^+\pi^+$ & $-0.017 \pm 0.042$ \\
\hline
\end{tabular}
\end{table}

Statistical errors in the first three modes in Table~\ref{tab:asy}
are more than two times smaller than in previous experiments, and the 
final mode represents a new measurement.  
It is apparent that the level of experimental uncertainty is still 
an order of magnitude larger than the predicted ${\cal O}(10^{-3})$ 
SM effects described earlier in Sect.~3.2. 
\section{FUTURE CHARM STUDIES}
Experimental advances will continue 
in the near term, such as completion of the E791 data analysis, 
the forthcoming Fermilab experiments 831 and 835, ongoing 
CLEO measurements, {\it etc}.  Let us consider some 
longer term prospects.  
\subsection{Tau-charm factory}
This topic is covered by the next speaker$^{\cite{Ch96}}$ 
in these proceedings.  The reader should also consult the incisive 
summary of $D$ physics possibilities at a $\tau cF$ 
by the Orsay group.$^{\cite{Le92}}$   A facility running at $\psi (3770)$ 
and $\psi (4160)$ will produce {\it coherent} 
$P$-wave $D^0{\bar D}^0$ pairs.  There already exists an abundant 
literature on this interesting quantum mechanical 
system, starting with the original Bigi-Sanda 
analysis$^{\cite{BiSa86}}$, including Liu's preprint$^{\cite{Li95}}$ 
and continuing up to the thorough exposition of Xing.$^{\cite{Xi97}}$ 
\subsection{Asymmetric B-factory}
Although the main objective of $B$-factory physics is 
detection of CPV in the $b$-quark sector, one should 
keep in mind the potential for doing $c$-quark studies as 
well.  For example, a $D^0 - {\bar D}^0$ mixing search could 
entail running at $\Psi (4S)$, using $D \to K\pi\pi$ as a tag 
and employing the semileptonic decay $D^0 \to K\ell\nu$ to 
probe mixing.  It has been estimated that for 
a machine luminousity of ${\cal L} \simeq 3 \times 
10^{33}~{\rm cm}^{-2}{\rm s.}^{-1}$ and a run time of 
$10^7$~s. would yield roughly $5\times 10^4$ $D$'s per 
year.$^{\cite{Bl96}}$  Utilizing the semileptonic mode would present a 
relatively clean environment for detection.  
\subsection{Hadroproduction}
There is potential for doing a hadroproduction 
experiment with an even larger sample than E791.  
As with E791, one could use a $D^* \to D \pi_{\rm soft}$ tag, 
and there would be the usual hadroproduction background problems.  

In relation to a new interaction point (`C0') under 
construction at Fermilab, an Expression of Interest has been 
filed by the CHARM2000 collaboration for the purpose of studying 
a variety of charm physics items (CPV, FCNC, $L$=$1$ charm mesons, 
{\it etc}).  A wire target would be inserted in the beam halo, 
and with a spectrometer of relatively modest cost and existing 
 detector/trigger/data-acquisition technologies, it is anticipated 
that a high statistics, high impact experiment could be run 
in either collider or fixed target mode.  For example, 
assuming a run of $10^7{\rm s}$ with an interaction rate of 
$10^6~{\rm int}/{\rm s}$, a charm cross section of the form 
\begin{equation}
{\sigma_{D^0 + {\bar D}^0}\over \sigma_{\rm inel}} \simeq 
(6.5 \pm 1.1)\times 10^{-4}~A^{0.29} \ \ ,
\end{equation}
an acceptance-times-efficiency of $(10 \pm 3)\%$ for the 
proposed detection conditions and $B_{D \to K\pi} \simeq 0.05$, 
one calculates 
\begin{eqnarray}
\lefteqn{n_{D\to K\pi} =} \nonumber \\
& & 10^7{\rm s} \cdot 10^6{{\rm int}\over {\rm s}} 
\cdot 6.5\times 10^{-4}A^{0.29} \cdot 0.1 \cdot 0.04  \nonumber \\
&\simeq& 2.6 \times 10^7 ~A^{0.29} \nonumber \\
&\simeq& 10^8 \ \ . \nonumber
\end{eqnarray}
The charm yield of $n_D \simeq 10^8$/yr for reconstructed $D\to K\pi$ 
would be substantial.  At this level of statistics, searches for 
CPT-violation might become meaningful.$^{\cite{Co95}}$
\subsection{HERA}
One should not ignore the potential at HERA for making 
contributions to charm studies.  Measured $eP$ 
charm-production cross sections together with an integrated luminousity of 
$250~{\rm pb}^{-1}$ implies a yield of $c{\bar c}$ pairs in excess of 
$10^8$.  For a detailed study, see Ref.~\cite{He92} as well as the 
recent summaries of Ref.~\cite{Gr96}.  
\section{CONCLUDING REMARKS}
The charm sector of the Standard Model proved to be 
a fruitful area for study in 1996, while leaving 
lots yet to be done.  The everpresent need for data 
will continue to be met by experimentalists, next year and 
beyond.  Theorists will continue to use 
charm hadrons for probing QCD dynamics (while awaiting 
the ultimate ascendency of lattice-QCD studies) and 
will also extend the catalog of new physics signals.  
\section{ACKNOWLEDGEMENTS}
We are happy to acknowledge the useful inputs of Sandip 
Pakvasa, Guy Blaylock, Dan Kaplan, Jeff Appel and Jo\~ao Soares 
to this review.

\end{document}